\documentclass[sigconf]{acmart}

\usepackage{booktabs} % For formal tables

\setcopyright{none}
\begin{document}
\title{Deploying Deep Ranking Models for Search Verticals}
%\titlenote{Produces the permission block, and  copyright information}
%\subtitle{Extended Abstract}
%subtitlenote{The full version of the author's guide is available as \texttt{acmart.pdf} document}

\settopmatter{printacmref=false}
\renewcommand\footnotetextcopyrightpermission[1]{} % removes footnote with conference information in first column
\pagestyle{plain} % removes running headers

\author{Rohan Ramanath$^{*}$, Gungor Polatkan$^{*}$, Liqin Xu, Harold Lee, Bo Hu, Shan Zhou}	
\affiliation{
	\institution{LinkedIn Corporation}
}
\email{{rramanath, gpolatkan, lxu, hlee, bohu, shzhou} @ linkedin.com}


\begin{abstract}
In this paper, we present an architecture executing a complex machine learning model such as a neural network capturing semantic similarity between a query and a document; and deploy to a real-world production system serving 500M+ users. We present the challenges that arise in a real-world system and how we solve them. We demonstrate that our architecture provides competitive modeling capability without any significant performance impact to the system in terms of latency. Our modular solution and insights can be used by other real-world search systems to realize and product-ionize recent gains in neural networks.
\end{abstract}

\maketitle

\section{Introduction}
\let\thefootnote\relax\footnote{~~~~ \\ $^*$ Equally contributing authors}
Capturing the semantic similarity between a query and a set of documents is a well studied problem in the Information Retrieval community \cite{Bollegala07, Rodriguez03, Mohler09}.

LinkedIn Talent Solutions (LTS) provides innovative tools to help talent searchers (e.g. recruiters, hiring managers and corporations) around the world become more successful at talent acquisition. One important challenge is to translate the criteria of a hiring position into a search query; the searcher has to understand which skills are typically required for a position, which companies are likely to have such candidates, which schools the candidates are most likely to graduate from, etc. Moreover, the knowledge varies over time. As a result, as indicated by the LinkedIn search log data \cite{Thuc16}, often multiple attempts are required to formulate a good query. To help the searcher, LTS search provides advanced targeting criteria called \textit{facets} (i.e. skills, schools, companies, titles and many more). The query can be entered as free text, a facet selection or the combination of the two. This results in queries where semantic interpretation and segmentation becomes important, e.g. in the query ``java'' or ``finance'' the searcher could be searching for a candidate whose title contains the word or someone who knows a skill represented by the word. Relying on exact term or attribute match in faceted search for ranking is sub-optimal. In this paper, we investigate a method to improve the solution to the matching and ranking problem rather than focus on the query formulation.

Latent semantic models are commonly used to map a noisy high dimensional query to a low-dimensional representation to make the matching problem tractable \cite{Deerwester90}.  We extend  latent semantic models with a deep structure by projecting queries and talent attributes into a shared low-dimensional space where the relevance of a talent given a query is readily computed as the distance between them. In this paper, we propose an architecture in which a neural network scoring a query-member pair is split into 3 semantic pieces such that each piece is scored on a separate system with its own characteristics. Additionally, we implement and experiment with one specific instance of this architecture in production, computing semantic similarity (used in a downstream learning to rank model \cite{Cao07}) using online  low-dimensional vector representations in a scalable way (being able to score millions of LinkedIn members)  without compromising system performance or site stability. It is important to note that many of the considerations of this approach is generalizable to any search vertical and conceptually the ideas generalize to applications beyond search. The rest of the paper is organized into the semantic details of the model (Sections \ref{section:modeling}, \ref{section:design}), details of the system design (Sections \ref{section:galene}, \ref{section:system}) and results in Section \ref{section:results}

\begin{figure}
	\includegraphics[width=0.6\columnwidth]{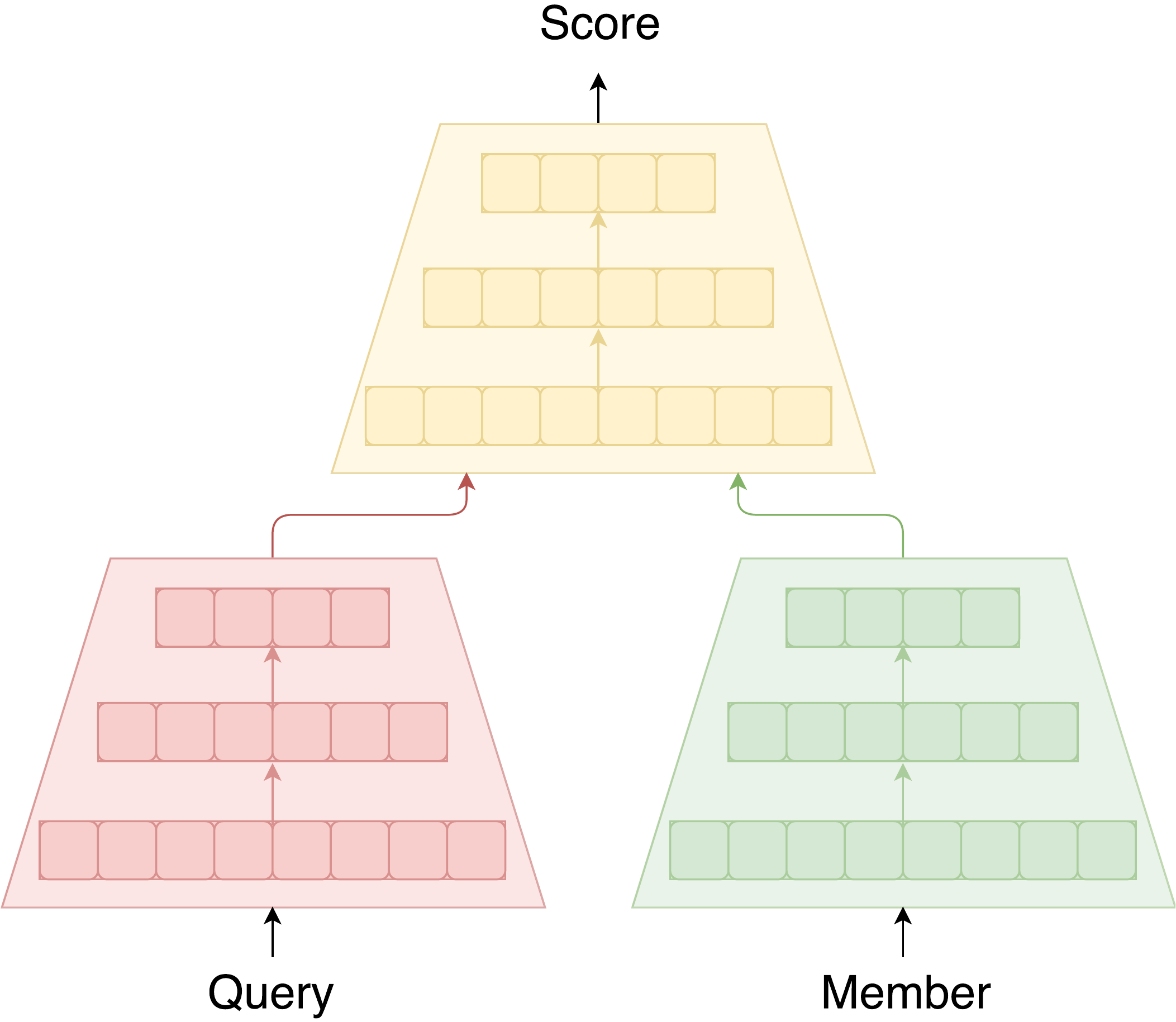}
	\caption{A typical siamese network with an additional crossing network. Although, this is trained as a single network using query log data, the architecture is split into $3$ semantic components during inference, which are implemented on separate physical systems}
    \vspace{-1.6em}
	\label{fig:model-design}
\end{figure}

\section{Modeling}
\label{section:modeling}
The problem of talent search can be formulated as follows; given a query \(q\) by a recruiter \(r\), rank a list of candidate LinkedIn members  \(m_1,m_2,...,m_d\) in decreasingly relevant order by learning a function (in this case a neural network scoring a query-member pair) , \(f: (q_i, r, m_j) \mapsto s_{i,j} \in \mathbb{R} \).  For the purposes of this paper, we make the model independent of the recruiter (\(r\)). Since our goal is to productionize this function, we study the characteristics of each system, and consequently each semantic piece (Figure \ref{fig:model-design}), required to serve a search result in Section \ref{section:system}.

In literature there have been different efforts trying to address similar problems. The models considered \cite{Huang13, Shen14, Hu14, Shan16} were point-wise methods with the focus on learning a function that scores the similarity between the query and a candidate(s). Our architecture in Figure \ref{fig:model-design} is a generalization of such models. In such a framework, the degree of model complexity of each module is dictated by 1) implementation and serving constraints, 2) requirements specified by a Service Level Agreement (SLA).

One drawback of the models in \cite{Huang13, Shen14, Hu14, Shan16} is that  they only consider text data. In LTS search, the query and talent are represented by multiple sources of data (profile picture, education, job history, skills and many more facets) and not just text. The problem of combining heterogeneous data of different modalities adds complexity to the ranking model.  In our experiments, we use the late crossing \cite{Shan16} variant of siamese networks \cite{Bromley94} since it allows us to compute scores within strict SLAs to be served online. 

\begin{figure}
	\includegraphics[width=1.0\columnwidth]{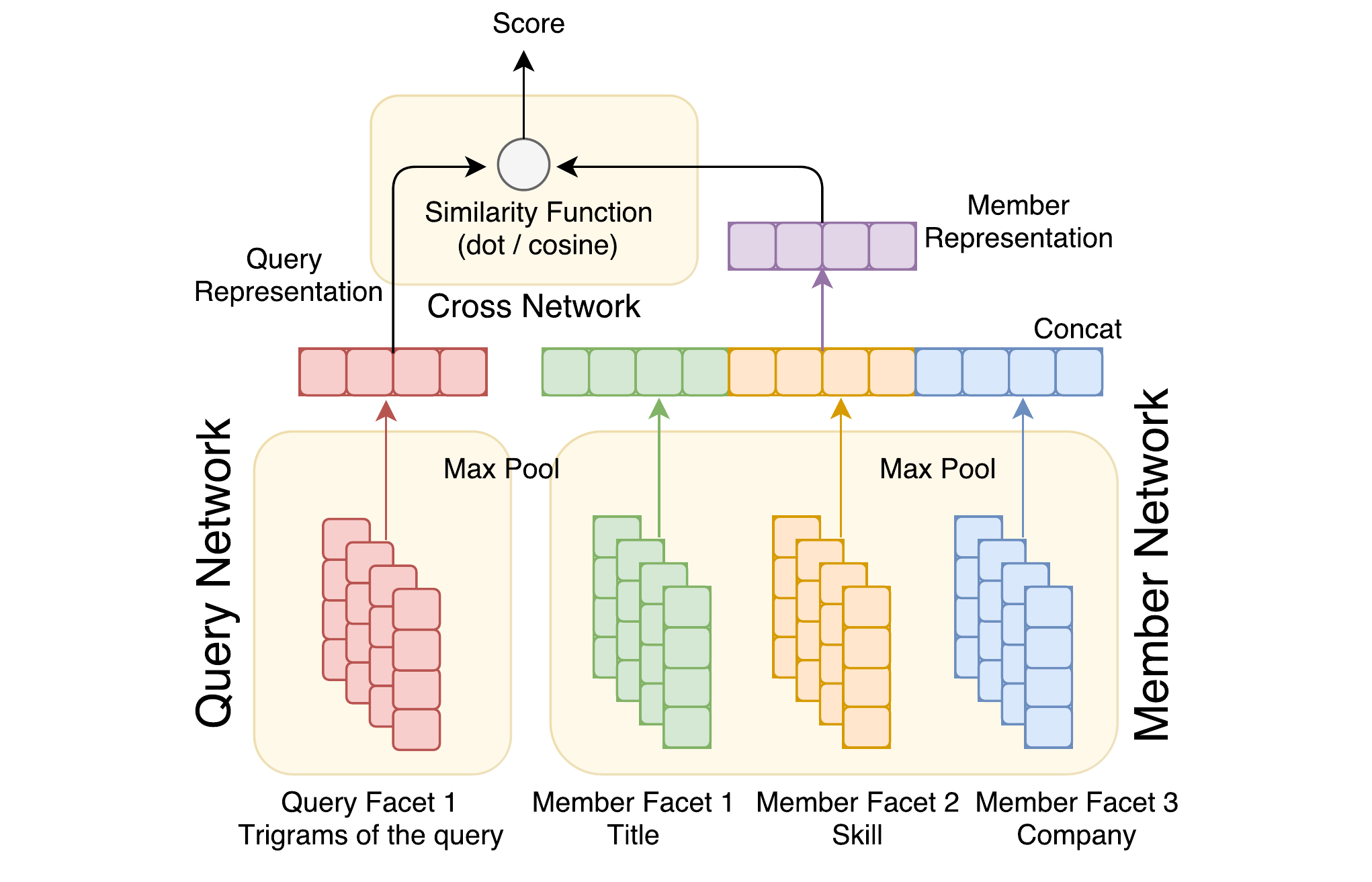}
	\caption{The two arm architecture with a shallow query arm (it is scored at real time) and a deep member arm. Note that the member arm is not only wider but also can be deeper.}
	\label{fig:model}
	\vspace{-2.5em}
\end{figure}

As shown in Figure \ref{fig:model}, the input to the model is a combination of text and facet attributes (A facet is any target-able attribute of the member). Each input layer converts the incoming attribute / text (ngram) from a list of categorical features to a single embedding (via pooling) and the aggregation layer simply stacks embeddings from multiple attributes to one  vector. Since the member arm has a richer source of input data, there is more opportunity to learn representative structures. This intuition manifests itself via a deeper and structurally richer (i.e. convolutions) member arm that eventually produces the member representation. The shorter query arm leverages query text and facets selected by the recruiter in the search UI to produce a query representation. The similarity layer (fully-connected, cosine, or any distance function) processes the query and member representations to produce a score that captures semantic similarity.

\vspace{-0.5em}
\section{Design Consideration}
\label{section:design}

We compute similarity \(sim(q, m)\) between a query \(q\) that contains terms \(\{t_1, t_2, \dots\}\) and member \(m\) that has attributes \(\{a_1, a_2, \dots \}\). The terms and member attributes could be keywords, tokens, or attributes of a user profile like skills, titles or company/school the user identifies with. Our approach is to use latent representations to compute similarity \(sim(q, m)\), but key question is what kinds of entities can we learn representations for? Would we use the representation of the entire query and  the entire member profile? Or, do we want to use representation of individual query term and member attribute? Depending on the entity representation chosen, we present two potential solution and their corresponding implementation details.

\subsection{Token Level Embeddings}

The first approach is to use the embedding vectors (i.e., latent representations) of query terms (i.e., tokens) \(\{t_1, t_2, \dots\}\) and member attributes (i.e., tokens) \(\{a_1, a_2, \dots\}\) to compute the query-member similarity. The token embeddings could be used to compute \(sim(q, m)\) in one of the following ways:

\begin{itemize}
	\item  Aggregating the similarity between individual query terms and member attributes \(sim(t_i, a_k)\).  Multiple aggregation strategies could be used and one such strategy is to add each similarity score as a feature to a linear model. The advantages of such as models are, (1) Easy path to productionization: Use an off-heap dictionary (or key-value store) containing the token embeddings in the online service, (2) No loss of information for tail queries or rare documents, since the information stored is at the token level. However, some disadvantages of this approach are that (1) The dictionary size has limitation because it is impractical to store more than a couple of hundreds of MB, (2) If the query contains a lot of terms, and member has a lot of attributes, computing similarities can be pretty time-consuming.
	\item Use a nonlinear function such as neural networks to get query-member similarity using the token level embeddings as features. The advantage of using nonlinearity is the richer set of interaction features that can be extracted from the raw data. However, as one stacks on layers in the network, the latency to score the function gets prohibitively expensive. The additional cost comes from the fact that for each query thousands of members need to be scored at run-time. The solution might be tractable if it were only a few query-member pairs that would need to be evaluated. To further this solution, it could be used as a feature in a downstream (i.e. broker) re-ranker that has significantly fewer query-member pairs as compared to the primary ranker in the search nodes.
\end{itemize}

It is worthy of mention that the above solution works in cases when the query member similarity can be decomposed into a function of individual term-attribute similarity. If that assumption does not hold, we cannot follow this approach. In cases when the vocabulary size of the tokens is too large, we cannot store the representation for all entities. In this case, we store representations for the top $K$ entities and this will decrease accuracy (or coverage) of our model. Additionally, if we deploy the vectors as a off-heap dictionary then versioning and testing multiple generations of embeddings is not a clean engineering solution because, (1) The size of the dictionary will grow at least linearly as the number of versions increase, (2) Deployment issues start to arise as the size of the deployable grows in a distributed multi-node deployment.

\subsection{Document Level Embeddings}

The second approach is to retrieve the representation (i.e., embedding) for the entire query and the member (i.e., document). This solution is applicable only when the query distribution has a short tail, i.e. the head queries serve a significant portion of the traffic. In such a situation one can learn a complex function to represent the query and the member and store the resulting query and member representations on key-value stores. One disadvantage of this approach is the limit of space and latency of storing such dense real-valued vectors in the forward index of a search engine. A workaround for other most search verticals could be to use an external key-value store to persist the member representation, but Galene's (Sec \ref{section:galene}) design restrict the search nodes from making external service calls.

\subsection{Hybrid Approach}

Since the query distribution for LTS search does not have a short tail, we could not use the document level embedding to pre-compute the query representation. Additionally, since the number of members that need to be scored for each query is of the order of a hundred thousand we cannot use token level representation for the member side because the memory and latency considerations are restrictive. Our solution involves using token level embeddings for one of the two sides (the query side) and document level embeddings for the other (the member side). We handle versioning by encoding version information into the field of the forward index.

\section{Galene}
\label{section:galene}

Although, our solution generalizes to any search vertical, we've integrated it into Galene for the purposes of this experiment. In this section, we introduce some components of Galene so that the reader could replicate our solution to other verticals. Galene is built on top of Apache Lucene, which is a widely used open-source search engine library for general purposes. Its major goal is to build a unified search engine system, providing the following (but not limited to) important additional features Lucene cannot support, by allowing users to:

\begin{figure}
	\includegraphics[width=0.8\columnwidth]{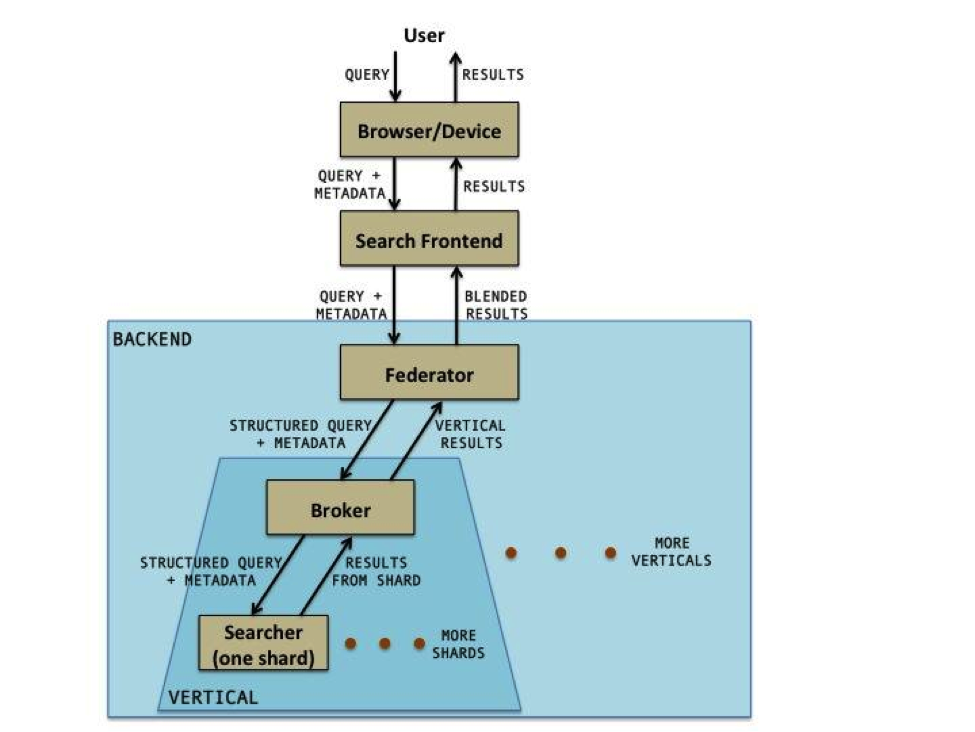}
	\caption{The life cycle of a query in Galene \cite{Sankar14}, the search infrastructure used at LinkedIn}
	\vspace{-1.75em}
	\label{fig:galene}
\end{figure}

\begin{itemize}
	\item Assign a globally unique identifier (UID) for each document, for example, memberId for the member profile. 
	\item Search over both offline indexes and real-time updates at the same time.  The offline indexes are built from our Hadoop Indexing pipeline periodically. The real-time updates are from Kafka. 
	\item Plug in any relevance functions and algorithms, bypassing the limited features available in Lucene's fixed scoring framework. Users can design their own relevance functions on a rich set of information about search hits, including term frequency, doc frequency, matched terms, and any metadata associated with the search hit document, et.c. 
\end{itemize}

Relevance modules can be plugged into Galene as new GaleneCollectors (which are extensions of Lucene collectors). Lucene collectors are primarily used to gather raw results from a search, and implement sorting or custom result filtering, collation, and so on. Similarly, GaleneCollectors can:

\begin{itemize}
	\item Collect the raw search results in a similar way as Lucene collector, plus additional early termination support.
	\item Collect the forward index, which could be anything. For example, it can be used to store all information you want to use to score and document.  
	\item Pluggable scoring mechanism. Users could treat it as a data provider (offering the information of a search hit, document info (forward index) or any other custom info (forward index) and apply any relevance functions on the data.
\end{itemize}

Sankar and Makhani \cite{Sankar14} give a good overview of the Galene stack and the life cycle of a search query is shown in Figure \ref{fig:galene}.
\section{System Details}
\label{section:system}

Our main design principle is to divide and conquer. Though implementations can be different, many practical search systems have three 3 main parts while  serving a query-member pair: 
\begin{itemize}
\item Offline distributed processing  (e.g. Hadoop, Spark)  to process offline data and lower the load on the online system in document processing and index preparation, 
\item Online query processing \cite{Betz13} for receiving the search request and performing an early evaluation and processing of the query, 
\item Searchers, the distributed platform carrying the index and performing the search based on the  processed query and previously prepared offline data. 
\end{itemize}
In our modularization of the model, we follow a similar pattern and divide the model in a similar way to the system. In our semantic split of the model, offline processing corresponds to the member network, online processing corresponds to the query network and searchers correspond to the cross network. Our implementation makes use of this pairing for executing and scoring of each piece of the model.

\subsection{Offline Processing} 
The offline distributed piece is used mostly for member network processing. Since the member profiles (education, job history, skills and many more facets) are known offline, we pre-compute the member representation using Tensorflow \cite{Abadi16} on Hadoop, compress and store this resultant vector in the forward index of the searcher. One can tolerate infrequent updates to the member representation because the member profile information is relatively static. Additionally, since the member representation is evaluated offline, we can experiment with more aggressive architectures (and depth) for the member arm.

\subsection{Online Query Processing} 
The online service is responsible for processing each search request. In this particular implementation we talk about a REST service. The query is evaluated and processed on the fly to extract query features like trigrams of text and search facets such as skill, title, company. The query network uses this as the input to produce a query representation as the output. Since the module is scored at real time and has tight SLAs, the network complexity is limited by the time to score. To simplify the discussion,  let us assume we just have one attribute, i.e. title $(t)$ on both the query and member side. Everything that follows can be easily extended to any number and types of attributes. A key-value store is used to store attribute, facet vectors, i.e. one vector for every title $t_i$ (or one vector for every \textit{ngram} if we consider \textit{ngrams} of the raw text as the query feature). The search frontend parses the query (the tagged textual query and selected facet) to determine all the titles targeted by the viewer.  The vectors corresponding to all the targeted titles are retrieved and the query arm of the network is evaluated in the search frontend. The resultant query representation is then inserted into the query meta data in the call to the search backend. Although evaluating the query arm can be computationally expensive (depending on the depth), this happens only once for a search request unlike the member arm of the network. An alternate solution could involve pre-computing and storing the query representation for the head queries and then directly retrieving them. However, further analysis of the query distribution did not reveal a power law, mostly because of the complexity introduced by facets and their interaction with the free-form text query.

\subsection{Searcher} 
\begin{figure}
	\includegraphics[width=\columnwidth]{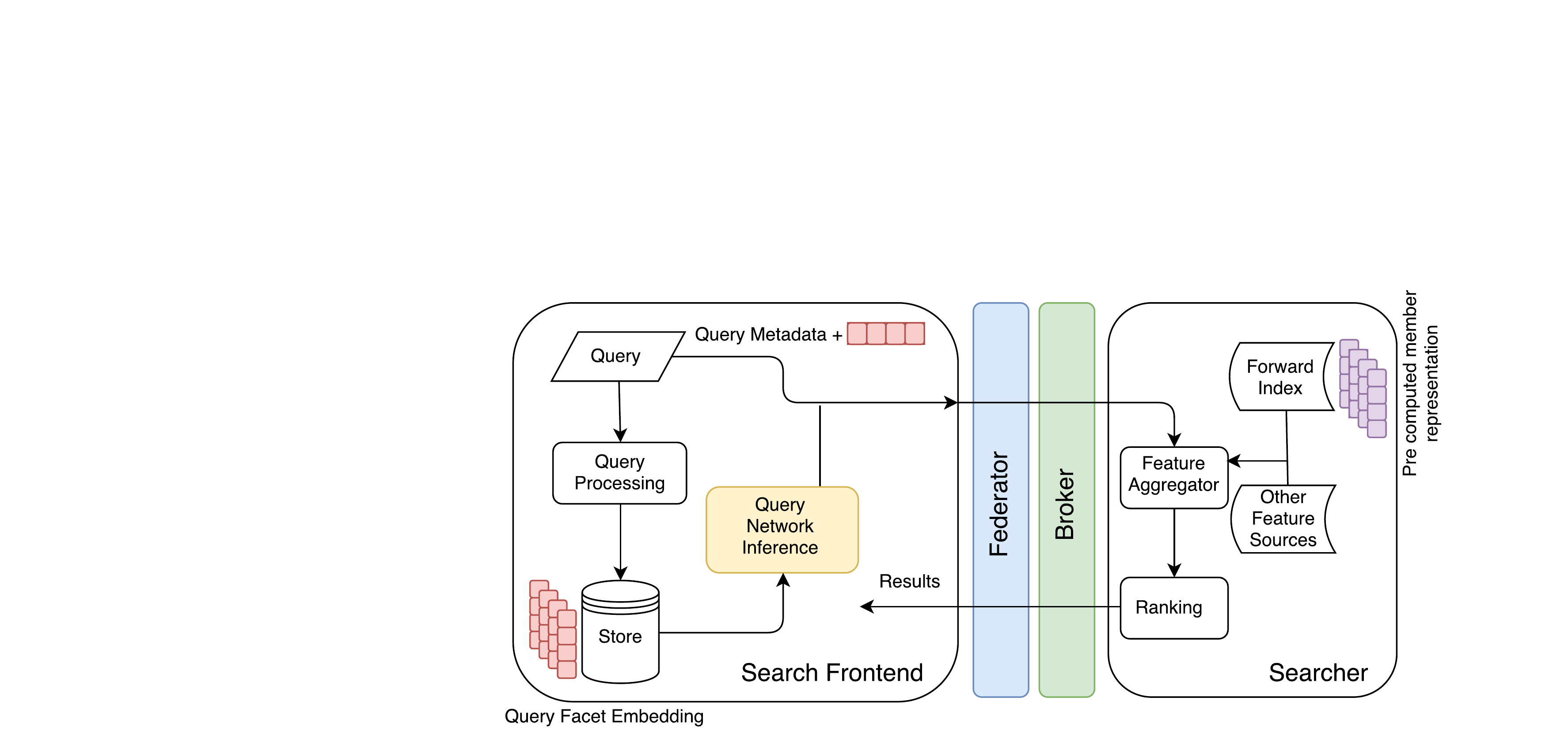}
	\caption{Implementation of the query and member similarity within the search infrastructure.}
	\label{fig:architecture}
\end{figure}

The third piece of the production pipeline is the LinkedIn search-as-a-service infrastructure, Galene,  \cite{Sankar14} for cross network as a final scoring. The offline generated member representation and REST service generated query representation are unified on the search nodes where the final piece of the scorer is evaluated. Galene is built over Lucene and most concepts discussed here will apply to other search frameworks. An important design decision in Galene that provides context to this work is that the backend (federator, broker and searcher in Figure \ref{fig:architecture}) should be self-sufficient and is not allowed to make external service calls. This design allowed for the search backend to be run against a suite of integration tests that evaluate the quality of the search index and ranking models before deployment. A side affect of this design is that it prevents one from using an external key-value store to store the pre-computed member representation. At request time, once the members have been retrieved for the query $q$, each member's representation (via the forward index) along with the query representation (via the request to the backend) are evaluated via the similarity layer in the searcher to produce a score for every query-member pair. This is then used as a feature in the ranking model.

\section{Results and Conclusion}
\label{section:results}

The system was ramped to $100\%$ of the LTS traffic and the feature generated by the network (one of the most important features in the model) is currently being used to rank search results. In the A/B test, we observed a 3\% increase in the overall precision which, per the domain and our experience, is a hard metric to move. In terms of system performance, there was no statistically significant difference in the latency (p50, p90, p99) of the search backend. Scoring the query network on the search frontend added $3 ms$ (p99) to the latency which was well within the SLA requirements since it needed to be computed only once per request. Our contribution can be summarized as, 
\begin{itemize} 
\item Demonstrate the use of neural network based embeddings to improve the relevance of search results, 
\item Propose an architecture that can be scored and leveraged by a real time production service, 
\item Show system scalability without any performance impact.
\end{itemize}

Deep Neural Networks provide very strong theoretical and experimental results in terms of accuracy. But it is often the engineering challenges blocking the realization of such models in production systems. This paper presents solutions to those challenges in real world production systems, which can be used to realize such gains.

\bibliographystyle{ACM-Reference-Format}
\bibliography{bibliography} 

\end{document}